\documentclass[12pt]{amsart}
\usepackage{amssymb}
\usepackage{color}
\usepackage{amsmath,epic,curves,amscd}
\usepackage[english]{babel}
\usepackage{graphicx}
\usepackage{comment}
\usepackage{appendix}
\usepackage{mathdots}
\usepackage[all]{xy}
\usepackage{relsize}
\usepackage{MnSymbol}
\DeclareMathSymbol{\shortminus}{\mathbin}{AMSa}{"39}

\theoremstyle{remark}

\begin{document}
\baselineskip 6.0 truemm
\parindent 1.5 true pc

\newcommand\lan{\langle}
\newcommand\ran{\rangle}
\newcommand\conv{{\text{\rm conv}}\,}

\newcommand\bfi{{\bf i}}
\newcommand\bfj{{\bf j}}
\newcommand\bfk{{\bf k}}
\newcommand\bfl{{\bf l}}
\newcommand{\comb}[2]{{}_{#1}\mathrm{C}_{#2}}

\newcommand\HS{{\text{\rm HS}}}
\newcommand\dist{{\text{\rm dist}}\,}
\newcommand\vol{{\text{\rm vol}}\,}
\newcommand\ghz{{\sf G}}
\newcommand\ghzst{{\text{\rm GHZ}}}
\newcommand\bisep{{\sf B}}
\newcommand\fbi{{\sf F}}
\newcommand\mermin{{\sf M}}
\newcommand\vlrd{{\text{\rm vlrd}}\,}
\newcommand\gen{{\ghz_n\setminus\bisep_n}}
\newcommand\xx{{\text{\sf X}}}

\title{Polytope structures for Greenberger-Horne-Zeilinger diagonal states}

\author{Kyung Hoon Han and Seung-Hyeok Kye}
\address{Kyung Hoon Han, Department of Data Science, The University of Suwon, Gyeonggi-do 445-743, Korea}
\email{kyunghoon.han at gmail.com}
\address{Seung-Hyeok Kye, Department of Mathematics and Institute of Mathematics, Seoul National University, Seoul 151-742, Korea}
\email{kye at snu.ac.kr}
\thanks{Both KHH and SHK were partially supported by NRF-2020R1A2C1A01004587, Korea}



\begin{abstract}
We explore the polytope structures for genuine entanglement,
biseparability, full biseparability and Bell inequality of
multi-qubit GHZ diagonal states. We first show that biseparable GHZ
diagonal states make hypersimplices inside the simplices consisting
of all GHZ diagonal states. Next, we consider full biseparability
which is equivalent to positive partial transpose for GHZ diagonal
states, and show that they make the convex hulls of simplices and
cubes. We also visualize which part of the simplex violates
multipartite Bell inequality. Finally, we compute precise volumes
for genuine entanglement, biseparability, full biseparability and
states violating Bell inequality among all GHZ diagonal states.
\end{abstract}

\maketitle

%%%%%%%%%%%%%%%%%%%%%%%%%%%%%%%%%%%%%%%%%%%%%%%%%%%%%%%%%%%%%%%%%%%%%%%%%%%%%%%%%%%%%%%%%%%%%%%%
%%%%%%%%%%%%%%%%%%%%%%%%%%%%%%%%%%%%%%%%%%%%%%%%%%%%%%%%%%%%%%%%%%%%%%%%%%%%%%%%%%%%%%%%%%%%%%%%
%%%%%%%%%%%%%%%%%%%%%%%%%%%%%%%%%%%%%%%%%%%%%%%%%%%%%%%%%%%%%%%%%%%%%%%%%%%%%%%%%%%%%%%%%%%%%%%%
%%%%%%%%%%%%%%%%%%%%%%%%%%%%%%%%%%%%%%%%%%%%%%%%%%%%%%%%%%%%%%%%%%%%%%%%%%%%%%%%%%%%%%%%%%%%%%%%
%%%%%%%%%%%%%%%%%%%%%%%%%%%%%%%%%%%%%%%%%%%%%%%%%%%%%%%%%%%%%%%%%%%%%%%%%%%%%%%%%%%%%%%%%%%%%%%%
%%%%%%%%%%%%%%%%%%%%%%%%%%%%%%%%%%%%%%%%%%%%%%%%%%%%%%%%%%%%%%%%%%%%%%%%%%%%%%%%%%%%%%%%%%%%%%%%
\section{Introduction}

The notion of entanglement arising from quantum mechanics is now recognized as
one of the most important resources in the current quantum information and computation theory.
The Greenberger-Horne-Zeilinger states \cite{{GHZ},{GHSZ}} are key examples of genuine entanglement
in multi-qubit systems, and have many applications in various fields of quantum information theory.
See survey articles \cite{{guhne_survey},{horo-survey}}.
They also play important roles in the classification of entanglement in multi-qubit systems
\cite{{dur},{dur_multi},{abls}}.

A mixed state is called {\sl separable} if it is a mixture of product states, and {\sl entangled}
if it is not separable. In the multi-partite systems, the notion of entanglement
depends on partitions of systems. A multi-partite state is called {\sl biseparable} if it
is a mixture of separable states with respect to bipartitions of systems, and called
{\sl genuinely entangled} if it is not biseparable. On the other hand, a state is called
{\sl fully biseparable} if it is biseparable with respect to any bipartitions of systems.

The GHZ diagonal states are mixtures of GHZ states \cite{{dur},{murao}}.
The mixture with the uniform distribution gives rise to the maximally mixed states,
that is, the scalar multiples of the identity.
By the results in \cite{{guhne10},{gao},{Rafsanjani},{han_kye_optimal}},
we have now complete criteria for biseparability and full biseperability of GHZ diagonal states.
We first note that those criteria are given by finitely many linear inequalities, and so the resulting
convex sets are polytopes.
We recall that a convex set in a finite dimensional space is called a {\sl polytope} if it has a finitely many extreme points.
It is well known that this is equivalent to the condition that it has finitely many facets, that is, maximal faces given by hyperplanes.
See \cite{{bron},{webster},{ziegler},{grun}} for examples.

The main purpose of this note is to explore the polytope structures for biseparable and fully biseparable GHZ diagonal states.
We recall \cite{han_kye_optimal} that a GHZ diagonal state is fully biseparable if and only if it is of PPT.
We also visualize which GHZ diagonal states violate Bell type inequalities.
We first note that the convex set $\ghz_n$ of all $n$-qubit GHZ diagonal states
is the regular simplex of dimension $d-1$ with the side length $\sqrt 2$, where we retain the notation $d=2^n$ throughout this note.
The GHZ states correspond to vertices and the maximally mixed state is located at the center of the simplex.
We show that the convex set $\bisep_n$ consisting of all biseparable GHZ diagonal states is
a {\sl truncation polytope} \cite{bron}, that is, a polytope obtained from a simplex by successive truncations of vertices.
Genuine entanglement among GHZ diagonal states are located in the truncated parts which consist of
$d$ pieces of $(d-1)$ regular simplices with the side length $\frac 1{\sqrt 2}$.
The remaining polytope $\bisep_n$ is the convex hull of midpoints of edges of $\ghz_n$, which is the half sized {\sl hypersimplex}
$\Delta_{d-1}(2)$ \cite{{ziegler},{grun}}.
On the other hand, the convex set $\fbi_n$ of all fully biseparable GHZ diagonal states is the convex hull
of the $(\frac d2-1)$ regular simplex and the $\frac d2$ regular cube which locate in the perpendicular position and share
only the maximally mixed state. We note that $\bisep_2=\fbi_2$ holds for the two qubit case, and their polytope structures are
already known in \cite{Lan_10}. We consider Mermin inequality as a multipartite Bell inequality, and
see that the part of $\ghz_n$ satisfying the inequality is also a truncation polytope by a single truncation.

With this information, we compute precise values of volumes, relative volumes and relative volume radii
for genuine entanglement, biseparability, full biseparability and violation of Mermin inequality among all GHZ diagonal states.
We also find the largest balls inside the polytopes $\ghz_n$, $\bisep_n$ and $\fbi_n$.
It is interesting to note that all of them coincide.

%%%%%%%%%%%%%%%%%%%%%%%%%%%%%%%%%%%%%%%%%%%%%%%%%%%%%%%%%%%%%%%%%%%%%%%%%%%%%%%%%%%%%%%%%%%%%%%%
%%%%%%%%%%%%%%%%%%%%%%%%%%%%%%%%%%%%%%%%%%%%%%%%%%%%%%%%%%%%%%%%%%%%%%%%%%%%%%%%%%%%%%%%%%%%%%%%
%%%%%%%%%%%%%%%%%%%%%%%%%%%%%%%%%%%%%%%%%%%%%%%%%%%%%%%%%%%%%%%%%%%%%%%%%%%%%%%%%%%%%%%%%%%%%%%%
%%%%%%%%%%%%%%%%%%%%%%%%%%%%%%%%%%%%%%%%%%%%%%%%%%%%%%%%%%%%%%%%%%%%%%%%%%%%%%%%%%%%%%%%%%%%%%%%
%%%%%%%%%%%%%%%%%%%%%%%%%%%%%%%%%%%%%%%%%%%%%%%%%%%%%%%%%%%%%%%%%%%%%%%%%%%%%%%%%%%%%%%%%%%%%%%%
%%%%%%%%%%%%%%%%%%%%%%%%%%%%%%%%%%%%%%%%%%%%%%%%%%%%%%%%%%%%%%%%%%%%%%%%%%%%%%%%%%%%%%%%%%%%%%%%
\section{Polytopes}

Throughout this note, we denote by $I_{n}$ the set of all {\sl $n$-bit indices} which are, by definition, functions from
$\{1,2,\dots,n\}$ into $\{0,1\}$. Therefore, they are \{0,1\} strings of length $n$.
For examples, we have
$I_2=\{00,01,10,11\}$, $I_3=\{000,001,010,011,100,101,110,111\}$,
and so, $I_n$ may be considered as the set of natural numbers from $0$ to $2^n-1$ with the binary expression.
For a given index $\bfi\in I_n$, the index $\bar\bfi\in I_n$ is defined by $\bar\bfi(k)=\bfi(k)+1$ mod $2$.
For an example, we have $\overline{010}=101$.

\subsection{GHZ diagonal states}
For each index $\bfi\in I_n$, the GHZ state is given by
$$
|\ghzst_\bfi\ran =\frac 1{\sqrt 2}(|\bfi\ran +(-1)^{\bfi(1)}|\bar\bfi\ran).
$$
For examples, we have
$|\ghzst_{00}\ran=\frac 1{\sqrt 2}(|00\ran+|11\ran)$ and
$|\ghzst_{101}\ran=\frac 1{\sqrt 2}(|101\ran-|010\ran)$.
An $n$ qubit state is called {\sl GHZ diagonal} if it is a convex combination
of the above states. We remind the readers of our convention $d=2^n$. Because the above states are orthonormal, we see that
the convex set $\ghz_n$ of all $n$-qubit GHZ diagonal states is the regular $(d-1)$ simplex with $d$ vertices
$$
v_\bfi:=|\ghzst_\bfi\ran\lan\ghzst_\bfi|,\qquad \bfi\in I_n,
$$
and every GHZ diagonal state is uniquely written by
$$
\varrho_p:=\sum_{\bfi\in I_n} p_\bfi v_\bfi
$$
with a probability distribution $p$ over $I_n$.
For a given fixed index $\bfi$, we note that the convex hull
$$
\ghz_n^\bfi:= \conv \{v_\bfj:\bfj\neq\bfi\}=\{\varrho_p\in \ghz_n: p_\bfi=0\}
$$
is a facet of $\ghz_n$ which is given by the hyperplane
$p_\bfi=0$, and every facet of $\ghz_n$ arises in this way. If we
endow the index set $I_n$ with the lexicographic order, then the GHZ
diagonal state $\varrho_p$ may be expressed by the following
$d\times d$ matrix
\begin{equation}\label{matrix}
\xx(a,z):=\left(
\begin{matrix}
a_{00\dots 0} &&&&&&&&& z_{00\dots 0}\\
& \ddots &&&&&&& \iddots &\\
&& a_{\bfi} &&&&& z_{\bfi} & \\
&&& \ddots &&& \iddots &&\\
%&&&& a_{01\dots 1}&z_{{01\dots 1}} &&&\\
%&&&&  z_{10\dots 0}&a_{10\dots 0}&&&\\
&&& \iddots &&& \ddots &&\\
&&  z_{\bar\bfi} &&&&& a_{\bar\bfi} &\\
& \iddots &&&&&&& \ddots &\\
 z_{11\dots 1} &&&&&&&&& a_{11\dots 1}
\end{matrix}
\right),
\end{equation}
with $a_\bfi=\frac 12(p_\bfi+p_{\bar\bfi})$ and $z_\bfi=\frac
{(-1)^{\bfi(1)}}2(p_\bfi-p_{\bar\bfi})$.

With the uniform distribution, we have the center point
$c:=\frac 1d\sum_\bfi v_\bfi=\frac 1d I_d$
of the simplex $\rm GHZ_n$, which is the maximally mixed state.
The center of the facet $\ghz_n^\bfi$ is given by
$c^\bfi:=\frac 1{d-1}\sum_{\bfj\neq\bfi}v_\bfj$.
We note that the three points $v_\bfi$, $c$ and $c^\bfi$ are collinear.
We also see that the \lq\lq height\rq\rq\ of the simplex $\ghz_n$ is
the distance between $v_\bfi$ and $c^\bfi$, which is given by
$\sqrt{d/(d-1)}$ with respect to the Hilbert-Schmidt norm.
The center point $c$ divides the height by
the ratio $1-\frac 1d:\frac 1d$, and so it approaches $c^\bfi$ as the number of qubit
increases. See {\sc Fig 1}.

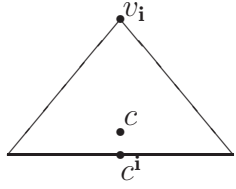
\begin{figure}
\begin{center}
\setlength{\unitlength}{1.5 truecm}
\begin{picture}(1.5,1.5)
\thinlines
\drawline(0,0)(1,1.2)(2,0)(0,0)
\put(1,1.2){\circle*{0.06}}
\put(1,0.2){\circle*{0.06}}
\put(1,0){\circle*{0.06}}
\put(1.02,1.23){$v_\bfi$}
\put(1.03,0.26){$c$}
\put(1,-0.22){$c^\bfi$}
\end{picture}
\end{center}
\caption{
The triangle and its base represent the simplex $\ghz_n$ and the facet $\ghz_n^\bfi$, respectively. The
center $c$ of the simplex approaches the center $c^\bfi$ of the facet $\ghz_n^\bfi$
as the dimension increases.
}
\end{figure}

\subsection{biseparable states}
It is known \cite{guhne10} that the GHZ diagonal state $\varrho_p$
of (\ref{matrix}) is biseparable if and only if the inequality
$|z_\bfi| \le \tfrac 12\sum_{\bfj\neq \bfi,\bar\bfi} a_\bfj$ holds for every $\bfi\in I_n$.
The inequality can be written as $|p_\bfi -
p_{\bar\bfi}| \le  \sum_{\bfj \ne \bfi, {\bar\bfi}} p_\bfj$, of which the right hand side is equal to $1-p_\bfi-p_{\bar\bfi}$.
Hence, $\varrho_p$ is biseparable if and only if
\begin{equation}\label{bisep}
p_\bfi\le \frac 12,\qquad {\text{\rm for every}}\ \bfi\in I_n.
\end{equation}
In other words, a GHZ diagonal state $\varrho_p$ is genuinely entangled if and only if $p_\bfi> \frac 12$
for some $\bfi\in I_n$. From this, we see that genuine entanglement is detected by the hyperplanes $p_\bfi=\frac 12$,
which also determine facets
$$
\bisep_{n}^{\bfi}:=\{\varrho_p\in \ghz_n: p_\bfi=\textstyle\frac 12\}
$$
of the convex set $\bisep_n$ consisting of all $n$-qubit biseparable GHZ diagonal states.
Especially, we see that $\bisep_n$ is a polytope; it has finitely many facets.
For each $\bfi\in I_n$, the region $\{\varrho_p\in\ghz_n: p_\bfi> \frac 12\}$ contains only one vertex $v_\bfi$,
and so $\bisep_n$ is a truncation polytope.
Genuine entanglement consists of such regions through indices $\bfi\in I_n$.
We note that an algebraic formula for genuinely multipartite concurrence for GHZ diagonal states
in \cite{Rafsanjani} is given by
$$
C_{\rm GM}(\varrho_p) = \max\{0, 2(\max_\bfi p_\bfi) -1\}.
$$
Therefore, we see that a level set of $C_{\rm GM}(\varrho)$ is parallel to a facet $\bisep_{n}^{\bfi}$ of
the convex set $\bisep_n$, and $C_{\rm GM}(\varrho)$ takes the maximum at vertices of $\ghz_n$.
%Thus, the GM concurrence of $\varrho$ is $c>0$ if and only if $\varrho$ belongs to the hyperplane $p_\bfi = (c+1) \slash 2$ for some $\bfi$.
%If we move the facets $p_\bfi = 1 \slash 2$ of the polytope $\bisep_n$ outward to the vertex $v_\bfi$,
%then we obtain the level sets of GM concurrence.

In order to understand the polytope structures of $\bisep_n$, we proceed to search for all extreme points. First of all,
we consider the case when $\varrho_p$ satisfies $p_\bfi=0$ or $p_\bfi=\frac 12$ for each $\bfi$.
In this case, there exist exactly two indices $\bfi,\bfj$ such that $p_\bfi=p_\bfj=\frac 12$,
and so we see that the resulting state
\begin{equation}\label{ext_bisep}
\varrho_p=\frac 12(v_\bfi+v_\bfj):= m_{\bfi,\bfj}
\end{equation}
is the midpoint of the edge of $\ghz_n$ connecting two
vertices $v_\bfi$ and $v_\bfj$. It is also clear that
$m_{\bfi,\bfj}$ is an extreme point of $\bisep_n$ since it is the
unique point of $\bisep_n$ on this edge by (\ref{bisep}). Note that
$m_{\bfi,\bar\bfi}$ is a diagonal state with two nonzero diagonal
entries. Conversely, suppose that $\varrho_p\in \bisep_n$ satisfies
$0<p_\bfi <\frac 12$ for some $\bfi$. Then we take the largest
$p_{\bfi_1}$ and the second largest $p_{\bfi_2}$, and consider the
line segment $\varrho_t=(1-t)m_{{\bfi_1},{\bfi_2}}+t\varrho_p$.
Since $p_{\bfi_1},p_{\bfi_2}>0$ and $p_\bfi<\frac 12$ for
$\bfi\neq{\bfi_1},{\bfi_2}$, we see that $\varrho_{1+\varepsilon}$
satisfies (\ref{bisep}) for small $\varepsilon>0$, and so
$\varrho_p$ is not an extreme point of $\bisep_n$. Therefore, we
conclude that the polytope $\bisep_n$ is the convex hull
of mid points of edges, as they are listed in
(\ref{ext_bisep}).
This also tells us that $\bisep_n$ is obtained by
maximal truncations of all vertices with the same size.
The polytope $\bisep_n$ can be considered as the half sized hypersimplex $\Delta_{d-1}(2)$,
whose vertex coordinates consist of $0$ and $1$, where the numbers of $0$ and $1$ are $(d-2)$ and $2$, respectively.
Note that the vertex coordinates of $\ghz_n$ also consists of $(d-1)$ number of $0$'s and one $1$.

We will bipartition extreme points into two groups. To do this,
we fix an index $\bfi$.
We note that an extreme point
$m_{\bfi,\bfj}$ belongs to the facet $\bisep_{n}^{\bfi}$ for every
$\bfj$ different from $\bfi$. If $\bfj,\bfk\neq\bfi$, then
$m_{\bfj,\bfk}$ belongs to the another facet
$$
\ghz_{n}^{\bfi}\cap\bisep_n=\{\varrho_p\in \bisep_n: p_\bfi=0\}
$$
of $\bisep_n$ which is determined by
the hyperplane $p_\bfi=0$. Therefore, we see that extreme points of
$\bisep_n$ are bipartitioned into two groups, one group in the facet $\bisep_n^\bfi$
and other group in the facet $\ghz_{n}^{\bfi}\cap\bisep_n$.
Therefore, we conclude that $\bisep_n$ is the convex hull of two parallel facets $\bisep_{n}^{\bfi}$ and
$\ghz_{n}^{\bfi}\cap\bisep_n$. We note that $\bisep_{n}^{\bfi}$ is
the $(d-2)$ simplex. On the other hand, $\ghz_{n}^{\bfi}\cap\bisep_n$ is the the half sized hypersimplex $\Delta_{d-2}(2)$
sitting in $\ghz_{n}^{\bfi}$. Every index $\bfi$ corresponds
to such a bipartion of extreme points, and corresponding two facets.
Therefore, the number of facets is given by $2d$. See {\sc Figure 2}
for $2$-qubit case. We finally note that every extreme point
$m_{\bfi,\bfj}$ is contained in exactly $d$ facets; $\bisep_n^\bfi$,
$\bisep_n^\bfj$ and $\ghz_n^\bfk\cap \bisep_n$ for
$\bfk\neq\bfi,\bfj$.

Our geometric approach also gives rise to a simple proof for
the characterization \cite{guhne10} of biseparability among GHZ diagonal states.
For the nontrivial part to prove that the condition
(\ref{bisep}) implies biseparability,  it is enough to show that extreme points $m_{\bfi,\bfj}$ are biseparable.
To do this, let $S$ and $T$ be the set of natural numbers $k=1,2,\dots,n$ such that $\bfi(k)=\bfj(k)$ and $\bfi(k)\neq\bfj(k)$,
respectively. Then it is easily seen that $m_{\bfi,\bfj}$ is separable with respect to the bipartition $S\sqcup T$
of systems, as in the two qubit case.

\begin{figure}
\begin{center}
\setlength{\unitlength}{3 truecm}
\begin{picture}(1.5,1.5)
\thinlines
\drawline(0.5,1.5)(0,0.5)(0.2,0)(1.4,0.3)(0.5,1.5)
\drawline(0.5,1.5)(0.2,0)
\drawline(0.25,1)(0.1,0.25)(0.35,0.75)(0.8,0.15)(0.95,0.9)
\dottedline{0.01}(0,0.5)(1.4,0.3)
\dottedline{0.01}(0.25,1)(0.7,0.4)(0.95,0.9)
{\linethickness{0.4mm}\dottedline{0.0001}(0.25,1)(0.35,0.75)(0.95,0.9)}
{\linethickness{0.4mm}\dottedline{0.02}(0.1,0.25)(0.8,0.15)(0.7,0.4)(0.1,0.25)}
{\linethickness{0.4mm}\dottedline{0.02}(0.25,1)(0.95,0.9)}
\put(0.25,1){\circle*{0.02}}
\put(0.35,0.75){\circle*{0.02}}
\put(0.95,0.9){\circle*{0.02}}
\put(0.1,0.25){\circle*{0.02}}
\put(0.8,0.15){\circle*{0.02}}
\put(0.7,0.4){\circle*{0.02}}
\put(0.51,1.5){$v_{00}$}
\put(1.41,0.29){$v_{11}$}
\put(-0.13,0.48){$v_{01}$}
\put(0.08,-0.04){$v_{10}$}
\put(-0.05,1){$m_{00,01}$}
\put(0.97,0.9){$m_{00,11}$}
\put(-0.21,0.25){$m_{01,10}$}
\put(0.8,0.09){$m_{10,11}$}
\end{picture}
\end{center}
\caption{
The polytope $\ghz_2$ has $4$ vertices $v_{00}$, $v_{01}$, $v_{10}$ and $v_{11}$ which make the regular
$3$ simplex. The polytope $\bisep_2$ of biseparable states has six vertices which are midpoints of edges of $\ghz_2$.
In this picture, we have the vertex $v_{00}$ on the top level, the facet $\bisep_2^{00}$ of $\bisep_2$ on the middle level and
the facet $\ghz_{2}^{00}\cap\bisep_2$ on the bottom level. The facet $\bisep_{2}^{00}$
is the regular $2$ simplex, and $\ghz_{2}^{00}\cap\bisep_2$
is the half sized hypersimplex $\Delta_2(2)$ sitting in the $2$ simplex $\ghz_{2}^{00}$.
}
\end{figure}
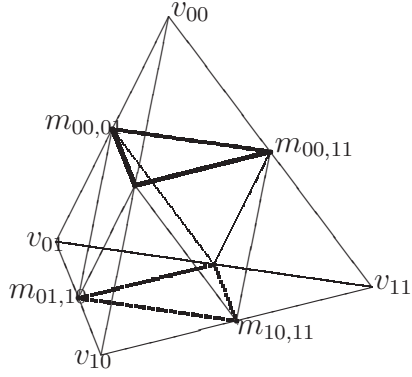

\subsection{fully biseparable states} Now, we turn our attention to full biseparability. It was
shown in \cite{han_kye_optimal} that a GHZ diagonal state is fully
biseparable if and only if it is of PPT with respect to any
bi-partition of parties. Therefore, a GHZ diagonal state $\varrho_p$
of (\ref{matrix}) is fully biseparable if and only if the following
\begin{equation}\label{fbi}
|z_\bfj| \le a_\bfi, \qquad
{\text{\rm for every}}\ \bfi,\bfj\in I_n
\end{equation}
holds. Note that the above inequalities are combinations of linear inequalities, and so the convex set
$\fbi_n$ of all fully biseparable GHZ diagonal states is also a polytope.

For a given $\varrho_p\in \fbi_n$ in (\ref{matrix}), we denote by $\lambda=\max|z_\bfi|$. Then we see that
$\varrho_p$ is the sum of the diagonal unnormalized state
$\sum_\bfi (a_\bfi-\lambda)|\bfi\ran\lan\bfi|$ and another unnormalized state $\xx({\lambda \bf 1},z)$ with the notation in (\ref{matrix}),
where ${\bf 1}_\bfi=1$ for each $\bfi\in I_n$. Since $a_\bfi=a_{\bar\bfi}$
and $m_{\bfi,\bar\bfi}=\frac 12(|\bfi\ran\lan\bfi|+|\bar\bfi\ran\lan\bar\bfi|)$, we have
$$
\varrho_p
=\sum_{\bfi\in I_n} (a_\bfi -\lambda) m_{\bfi,\bar\bfi}+\lambda d \, \xx(\textstyle\frac 1d {\bf 1},w)
$$
with $w=\frac 1{d} \frac z{\lambda}$ satisfying $-\frac 1d\le w_\bfi\le\frac 1d$ and $w_\bfi = w_{\bar\bfi}$ for each index $\bfi\in I_n$.
Therefore, we see that $\fbi_n$ is the convex hull of the following two polytopes
$$
\begin{aligned}
\fbi_n^\triangle
&:=\conv\{m_{\bfi,\bar\bfi}: \bfi\in I_n\},\\
\fbi_n^\square
&:=\{\xx(\textstyle\frac 1d {\bf 1},w):-{1 \over d} \le w_\bfi \le {1 \over d},\ w_\bfi = w_{\bar\bfi}\}.
\end{aligned}
$$
We note that $\fbi_n^\triangle$ consisting of diagonal states
is the regular $(\frac d2-1)$ simplex with the unit side length since $\{m_{\bfi,\bar\bfi}\}$ is an orthogonal family
with the uniform norm $1/{\sqrt 2}$.
It is clear that $\fbi_n^\square$ is the regular $\frac d2$ cube with the side length $2\sqrt 2/d$.

Suppose that $\sigma$ is a collection of $2^{n-1}$ indices
which has exactly one index among $\bfi$ and $\bar\bfi$. In case of two qubit, we have
four such choices; $\{00,01\}$, $\{00,10\}$, $\{11,01\}$ and $\{11,10\}$. In general, we have $2^{d/2}$ choices for the $n$-qubit case. We denote by
$v_\sigma^\square$ the GHZ diagonal state with the uniform distribution over $\sigma$, that is, we define
\begin{equation}\label{ext_fbi}
v_\sigma^\square:=\frac 2d\sum_{\bfi\in\sigma} v_\bfi=\xx(\textstyle\frac 1d{\bf 1},w),
\end{equation}
where $w_\bfi=w_{\bar\bfi}=(-1)^{\bfi(1)}\frac 1d$ for $\bfi\in\sigma$.
These states $v_\sigma^\square$'s are vertices of the cube $\fbi_n^\square$.
In fact, $v_\sigma^\square$ is an extreme point of
$\fbi_n$  since it is the only one point of $\fbi_n$
in the face $\ghz_n^\sigma$ of $\ghz_n$ generated by $v_\bfi$ with $\bfi\in\sigma$ by the PPT condition.
Therefore, there are exactly $\frac d2+2^{d/2}$ extreme points of $\fbi_n$.
Note that extreme points of $\fbi_3$ together with $\bisep_3$ have been found in
\cite{han_kye_pe,han_kye_lattice_id}.

In conclusion, the polytope $\fbi_n$ of all fully biseparable GHZ diagonal states
is the convex hull of the regular $(\frac d2 -1)$ simplex $\fbi_n^\triangle$ and the regular
$\frac d2$ cube $\fbi_n^\square$. Two polytopes $\fbi_n^\triangle$ and $\fbi_n^\square$
are perpendicular, and share only one point which is the maximally mixed state.
We see by \cite[Proposition 3.1]{kye_05_decom} that every face of $\fbi_n$ is the convex hull of a (possibly empty)
face of $\fbi_n^\triangle$ and a (possibly empty) face of $\fbi_n^\square$.
Because both $\fbi_n^\triangle$ and $\fbi_n^\square$ contain the maximally mixed state which is an interior point
of $\fbi_n$, we see that every facet of $\fbi_n$ is given by the convex hull of proper faces of $\fbi_n^\triangle$ and $\fbi_n^\square$.
On the other hand, facets of $\fbi_n^\triangle$ and $\fbi_n^\square$ are given by
\begin{equation}\label{two_facets}
\conv\{m_{\bfk,\bar\bfk} : \bfk \ne \bfi,{\bar\bfi}\}
\quad {\text{\rm and}}\quad
\{\xx(\textstyle\frac 1d {\bf 1},w) \in \fbi_n^\square : w_\bfj = (-1)^{\bfj(1)} \frac 1d\},
\end{equation}
for choices of indices $\bfi$ and $\bfj$, respectively, and their convex hull is the collection of
$\xx(a,w) \in \fbi_n$ satisfying $a_\bfi= (-1)^{\bfj(1)}w_\bfj $ determined by the identity in (\ref{fbi}).
Therefore, we conclude that the convex hull of facets of $\fbi_n^\triangle$ and
$\fbi_n^\square$ is a facet of $\fbi_n$, and every facet of $\fbi_n$ arises in this way.
We note that facets of $\fbi_n^\triangle$ and $\fbi_n^\square$ are determined by
choices of $\{\bfi,\bar\bfi\}$ and $\bfj$, respectively. They give rise to the facet
$\fbi_n^{\bfi,\bfj}$
of the polytope $\fbi_n$, which is the convex hull of two convex sets in (\ref{two_facets}). This facet is also given by the equation
\begin{equation}\label{cccc}
\fbi_n^{\bfi,\bfj}=\{\varrho_p\in \fbi_n: p_\bfi + p_{\bar\bfi} = p_\bfj-p_{\bar\bfj}\}
\end{equation}
in terms of probability distribution by (\ref{fbi}).
We also note that the number of facets of the polytope $\fbi_n$ is given by
$\frac{d^2}2$. See {\sc Figure 3} for the two qubit case.

\begin{figure}
\begin{center}
\setlength{\unitlength}{3 truecm}
\begin{picture}(1.5,1.5)
\thinlines
\drawline(0.5,1.5)(0,0.5)(0.2,0)(1.4,0.3)(0.5,1.5)
\drawline(0.5,1.5)(0.2,0)
\drawline(0.25,1)(0.35,0.75)(0.95,0.9)
\drawline(0.25,1)(0.1,0.25)(0.35,0.75)(0.8,0.15)(0.95,0.9)
\dottedline{0.01}(0,0.5)(1.4,0.3)
\dottedline{0.01}(0.1,0.25)(0.8,0.15)(0.7,0.4)(0.1,0.25)
\dottedline{0.01}(0.25,1)(0.7,0.4)(0.95,0.9)
\dottedline{0.01}(0.25,1)(0.95,0.9)
{\linethickness{0.4mm}\dottedline{0.001}(0.25,1)(0.35,0.75)(0.8,0.15)}
{\linethickness{0.4mm}\dottedline{0.02}(0.25,1)(0.7,0.4)(0.8,0.15)}
{\linethickness{0.6mm}\dottedline{0.02}(0.95,0.9)(0.1,0.25)}
\put(0.25,1){\circle*{0.02}}
\put(0.35,0.75){\circle*{0.02}}
\put(0.95,0.9){\circle*{0.02}}
\put(0.1,0.25){\circle*{0.02}}
\put(0.8,0.15){\circle*{0.02}}
\put(0.7,0.4){\circle*{0.02}}
\put(0.525,0.575){\circle*{0.04}}
\put(0.51,1.5){$v_{00}$}
\put(1.41,0.29){$v_{11}$}
\put(-0.13,0.48){$v_{01}$}
\put(0.08,-0.04){$v_{10}$}
\put(-0.05,1){$m_{00,01}$}
\put(0.97,0.9){$m_{00,11}$}
\put(-0.21,0.25){$m_{01,10}$}
\put(0.8,0.09){$m_{10,11}$}
\end{picture}
\end{center}
\caption{The polytope $\fbi_2$ is the convex hull of the line segment $[m_{00,11}, m_{01,10}]=F_2^\triangle$
and the square $F_2^\square$ which is perpendicular to the line segment. The square is the convex hull of $v_\sigma^\square$'s
with $\sigma$ among $\{00,01\}$, $\{00,10\}$, $\{11,01\}$ and $\{11,10\}$.
}
\end{figure}
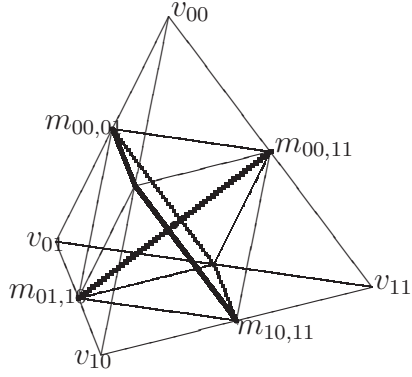

Using the geometry, we may also give a simple proof of the equivalence between PPT and full biseparability for GHZ diagonal states.
For the nontrivial part to show every GHZ diagonal state of PPT is fully biseparable,
it suffices to consider an extreme point $v_\sigma^\square$ of $\fbi_n$.
We fix a bipartition $S \sqcup T$, and denote by $\bar\bfi^S$
the index obtained by changing $k$-th symbols for $k\in S$, and similarly for $\bar\bfi^T$.
Then, for each $\bfi \in \sigma$, either $\bar{\bfi}^S \in \sigma$ or $\bar{\bfi}^T \in \sigma$ holds.
Thus, $v_\sigma^\square$ is the average of states of the form $m_{\bfi,\bar{\bfi}^S}$ or  $m_{\bfi,\bar{\bfi}^T}$ by (\ref{ext_fbi}).
For example, if $\sigma=\{000,001,011,101\}$ and the bi-partition is $A$-$BC$, then $v_\sigma^\square$ is the average of $m_{000,011}$ and $m_{001,101}$ which are $A$-$BC$ separable as in the two qubit case.

\subsection{Bell inequalities}
In this subsection, we consider multipartite Bell inequalities and figure out which parts violate the inequalities. See \cite{Belinskii} for a survey on Bell inequalities.
We begin with the Mermin inequality \cite{Mermin}, which considers two settings on each side.
Following \cite{guhne_survey}, we put
$$
M_n := \sum_\pi X_1X_2X_3X_4X_5 \dots X_n - \sum_\pi Y_1Y_2X_3X_4X_5 \dots X_n  + \sum_\pi Y_1Y_2Y_3Y_4X_5 \dots X_n - \cdots,
$$
where $X_i$ and $Y_i$ represent the Pauli matrices $\sigma_x, \sigma_y$ on the $i$-th qubit,
and $\sum_\pi$ represents the sum of all possible permutations of the qubits that give distinct terms.
Then the Mermin inequality is given by
$$
\lan M_n\ran_\varrho:={\text{\rm Tr}}(M_n\varrho)\le \mu_n:
=\begin{cases}
2^{n \slash 2}, &\quad n\ {\text{\rm even}},\\
2^{(n-1) \slash 2}, &\quad n\ {\text{\rm odd}}.
\end{cases}
$$

Recall the notation ${\bf 1}$ given by ${\bf 1}_\bfi=1$ for each $\bfi\in I_n$. We also use the notation ${\bf 0}$ given
by ${\bf 0}_\bfi=0$ for each $\bfi\in I_n$.
We have
$$
\begin{aligned}
M_n|{\bf 0}\rangle &= |{\bf 1}\rangle -\sum_\pi i^2|{\bf 1}\rangle +\sum_\pi i^4|{\bf 1}\rangle - \sum_\pi i^6|{\bf 1}\rangle + \cdots \\
&= \left(\comb{n}{0}+\comb{n}{2}+\comb{n}{4}+\comb{n}{6}+\cdots~\right)|{\bf 1}\rangle
= 2^{n-1}|{\bf 1}\rangle
\end{aligned}
$$
and $M_n|{\bf 1}\rangle = 2^{n-1}|{\bf 0}\rangle$ similarly.
Since $\{{1 \over \sqrt{2}}\sigma_x, {1 \over \sqrt{2}}\sigma_y\}$ is orthonormal, we have
$$
\|M_n\|_2^2 = 2^n \|({\textstyle\frac 1{ \sqrt{2}}})^n M_n\|_2^2 = 2^n(\comb{n}{0}+\comb{n}{2}+\comb{n}{4}+\comb{n}{6}+\cdots) = 2^{2n-1}.
$$
On the other hand, we also have
$$
|\langle{\bf 0}| M_n |{\bf 1}\rangle|^2 + |\langle{\bf 1}| M_n |{\bf 0}\rangle|^2=2^{2n-1}=\|M_n\|_2^2,
$$
which implies $\langle \bfi| M_n |\bfj\rangle =0$ whenever $\{\bfi,\bfj\} \neq \{{\bf 0},{\bf 1}\}$.
Therefore, we have
$$
\langle \bfi| M_n |\bfj\rangle=\begin{cases}
2^{n-1},&\quad (\bfi,\bfj)=({\bf 0},{\bf 1}), ({\bf 1},{\bf 0}), \\
0&\quad {\text{\rm otherwise}}\end{cases}
$$
and so it follows that
$$
\lan M_n\ran_{\varrho_p}= 2^{n-1} (p_{\bf 0}-p_{\bf 1}),
$$
for a GHZ diagonal state $\varrho_p$.

Now, we conclude that a GHZ diagonal state $\varrho_p$ violates the Mermin inequality if and only if
$$
p_{\bf 0}-p_{\bf 1} > \nu_n:=
\begin{cases}
2/\sqrt d,&\quad n\ {\text{\rm is even}},\\
\sqrt 2/\sqrt d,&\quad n\ {\text{\rm is odd}}
\end{cases}
$$
and the GHZ state $\varrho_{\bf 0}$ violates the inequality maximally.
The hyperplane
$$
H_M:=\{\varrho_p\in \ghz_n: p_{\bf 0}-p_{\bf 1} = \nu_n\}
$$
is perpendicular to the edge $\overline{v_{\bf 0}v_{\bf 1}}$ of the simplex $\ghz_n$ of all GHZ diagonal states, and
meets the edges $\overline{v_{\bf 0}v_{\bfi}}$ for $\bfi\neq\bf 0$ at the points
\begin{equation}\label{points_mermin}
w_{\bf 1}:=\textstyle (\frac 12 +\frac 12 \nu_n)v_{\bf 0}+(\frac 12 -\frac 12 \nu_n)v_{\bf 1},\qquad
w_{\bfi}:=\nu_nv_{\bf 0} +(1-\nu_n) v_{\bfi},\quad \bfi\neq {\bf 0},{\bf 1}.
\end{equation}
We note that $\nu_n=\frac 12$ for $n=3,4$ and $\nu_n<\frac 12$ for $n\ge 5$. This means that the hyperplane $H_M$
is tangent to the facet $\bisep_n^{\bf 0}$ of the convex set $\bisep_n$ for $n=3,4$. Therefore, we see that
three or four qubit biseparable GHZ diagonal states never violate the Mermin inequality. On the other hand,
there exists $n$ qubit biseparable GHZ diagonal states which violate the inequality for $n\ge 5$.
See {\sc Figure 4}.

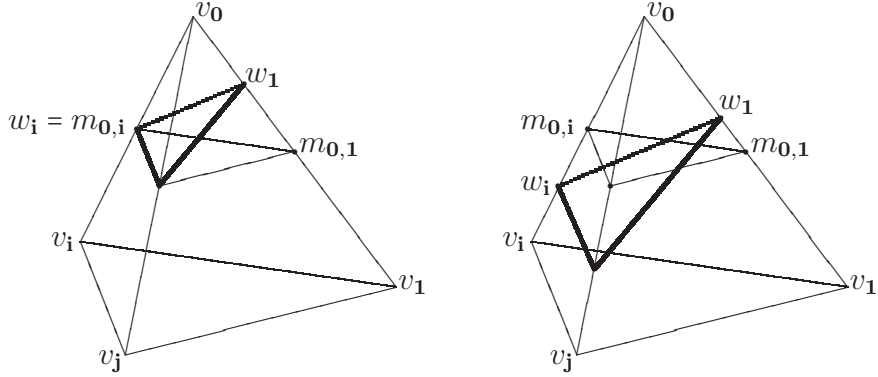
\begin{figure}
\begin{center}
\setlength{\unitlength}{3 truecm}
\begin{picture}(3.5,1.5)
\thinlines
\drawline(0.5,1.5)(0,0.5)(0.2,0)(1.4,0.3)(0.5,1.5)
\drawline(0.5,1.5)(0.2,0)
\dottedline{0.01}(0,0.5)(1.4,0.3)
\drawline(0.25,1)(0.35,0.75)(0.95,0.9)
\dottedline{0.01}(0.25,1)(0.95,0.9)
\put(0.25,1){\circle*{0.02}}
\put(0.35,0.75){\circle*{0.02}}
\put(0.95,0.9){\circle*{0.02}}
\put(0.51,1.5){$v_{\bf 0}$}
\put(1.41,0.29){$v_{\bf 1}$}
\put(-0.13,0.48){$v_{\bfi}$}
\put(0.08,-0.04){$v_{\bfj}$}
\put(-0.32,1){$w_\bfi=m_{{\bf 0},\bfi}$}
\put(0.97,0.9){$m_{{\bf 0},{\bf 1}}$}
%%%%%%%%%%%%%%%
{\linethickness{0.4mm}\dottedline{0.0001}(0.25,1)(0.35,0.75)}
{\linethickness{0.4mm}\dottedline{0.001}(0.35,0.75)(0.725,1.2)}
{\linethickness{0.4mm}\dottedline{0.02}(0.725,1.2)(0.25,1)}
\put(0.725,1.2){\circle*{0.02}}
\put(0.73,1.21){$w_{\bf 1}$}
%%%%%%%%%%%%%%%
\drawline(2.5,1.5)(2,0.5)(2.2,0)(3.4,0.3)(2.5,1.5)
\drawline(2.5,1.5)(2.2,0)
\dottedline{0.01}(2,0.5)(3.4,0.3)
\drawline(2.25,1)(2.35,0.75)(2.95,0.9)
\dottedline{0.01}(2.25,1)(2.95,0.9)
\put(2.25,1){\circle*{0.02}}
\put(2.35,0.75){\circle*{0.02}}
\put(2.95,0.9){\circle*{0.02}}
\put(2.51,1.5){$v_{\bf 0}$}
\put(3.41,0.29){$v_{\bf 1}$}
\put(1.87,0.48){$v_{\bfi}$}
\put(2.08,-0.04){$v_{\bfj}$}
\put(1.95,1){$m_{{\bf 0},\bfi}$}
\put(2.97,0.9){$m_{{\bf 0},{\bf 1}}$}
%%%%%%%%%%%%%%%
\put(2.8375,1.05){\circle*{0.02}}
\put(2.1175,0.746842105){\circle*{0.02}}
\put(2.2775,0.378){\circle*{0.02}}
{\linethickness{0.4mm}\dottedline{0.02}(2.8375,1.05)(2.1175,0.746842105)}
{\linethickness{0.4mm}\dottedline{0.001}(2.2775,0.378)(2.1175,0.746842105)}
{\linethickness{0.4mm}\dottedline{0.001}(2.2775,0.378)(2.8375,1.05)}
\put(2.83,1.09){$w_{\bf 1}$}
\put(1.95,0.73){$w_{\bfi}$}
\end{picture}
\end{center}
\caption{
The left and right tetrahedra represent the simplices $\ghz_n$ of all GHZ diagonal states for $n=3,4$ and $n\ge 5$, respectively.
Triangles with thick boundaries represent the hyperplane $H_M$, and the other triangles represent the facet $\bisep_n^{\bf 0}$
of the convex sets $\bisep_n$ of biseparable states.
The upper parts over the hyperplane $H_M$ consist of GHZ diagonal states which violate Mermin inequality.
}
\end{figure}

As for fully biseparable states or equivalently PPT states, we see that $\lan M_n\ran_{m_{\bfi,\bar\bfi}}=0$ and
$\lan M_n\ran_{v_\sigma^\square}=\pm 1$ for extreme points of $\fbi_n$. Therefore,
we see that no GHZ diagonal state of PPT violates the Mermin inequality.
Using the Lagrange method, the distance from the hyperplane $H_M$ to the convex set $\fbi_n$ is calculated by
$$
{\rm dist}(H_M, \fbi_n) = {1 \over \sqrt{2}}\left(\nu_n - {2 \over d}\right).
$$

%The exactly same argument may be applied for Ardehali inequality \cite{Ardehali,guhne_survey}.
We also consider the Ardehali inequality \cite{Ardehali}, which is another multi-partite Bell inequality.
The exactly same argument may be applied for Ardehali inequality in \cite{guhne_survey},
to see that the hyperplane determining the violation of Ardehali inequality is a translation of $H_M$.
In this case, we also see that this hyperplane meets the interior of $\bisep_n$ when and only when $n\ge 4$.

%%%%%%%%%%%%%%%%%%%%%%%%%%%%%%%%%%%%%%%%%%%%%%%%%%%%%%%%%%%%%%%%%%%%%%%%%%%%%%%%%%%%%%%%%%%%%%%%
%%%%%%%%%%%%%%%%%%%%%%%%%%%%%%%%%%%%%%%%%%%%%%%%%%%%%%%%%%%%%%%%%%%%%%%%%%%%%%%%%%%%%%%%%%%%%%%%
%%%%%%%%%%%%%%%%%%%%%%%%%%%%%%%%%%%%%%%%%%%%%%%%%%%%%%%%%%%%%%%%%%%%%%%%%%%%%%%%%%%%%%%%%%%%%%%%
%%%%%%%%%%%%%%%%%%%%%%%%%%%%%%%%%%%%%%%%%%%%%%%%%%%%%%%%%%%%%%%%%%%%%%%%%%%%%%%%%%%%%%%%%%%%%%%%
%%%%%%%%%%%%%%%%%%%%%%%%%%%%%%%%%%%%%%%%%%%%%%%%%%%%%%%%%%%%%%%%%%%%%%%%%%%%%%%%%%%%%%%%%%%%%%%%
%%%%%%%%%%%%%%%%%%%%%%%%%%%%%%%%%%%%%%%%%%%%%%%%%%%%%%%%%%%%%%%%%%%%%%%%%%%%%%%%%%%%%%%%%%%%%%%%
\section{Volume}

We note that the whole $n$-qubit GHZ diagonal states are trisected by the following three parts:
\smallskip

$\bullet$ $\ghz_n\setminus \bisep_n$: genuine entanglement,

$\bullet$ $\bisep_n\setminus\fbi_n$: biseparable but not fully biseparable states,

$\bullet$ $\fbi_n$: fully biseparable states.
\smallskip

\noindent
We first compute precise volumes for the above parts with respect to the
Hilbert Schmidt norm. We note that there are lots of estimates for the volumes of
separable states in various situations in the literature. See
\cite{{zyczk_03},{szarek_05},{slater_05},{aubrun_06},{Grabowski_13},{Singh_14},{Lancien_15}} for examples.

When two convex sets $C_1$ and $C_2$ with a common point are perpendicular to each other, we denote
by $C_1\diamondplus C_2$ the convex hull of them.
Since they are perpendicular, the common point is unique.
When $\Delta_p$ is the regular $p$ simplex with the side length $\ell$
and $C$ is a $q$-dimensional convex body with volume $V_0$, we will compute the volume $V_p$ of the convex set $\Delta_p\diamondplus C$.
When $p=1$, the volume $V_1$ of $\Delta_1\diamondplus C$ is given by
$V_1=\frac 1{1+q}\cdot\ell\cdot V_0$.
We also note that $\Delta_{p}=\left[h_p\Delta_1\right]\diamondplus\Delta_{p-1}$, where
$h_p=\frac 1{\sqrt{2}}\sqrt{\frac{p+1}{p}}$ is the \lq height\rq\ of the $p$ simplex with the unit side length.
We translate $C$ so that it meets $\Delta_{p-1}$.
Since $\Delta_p$ and $C$ are perpendicular, the volume $V_p$ does not change.
Therefore, we have the following inductive formula
$$
V_p
=\vol(\left[h_p\Delta_1\right]\diamondplus(\Delta_{p-1}\diamondplus C))
=\frac 1{p+q}\cdot {h_p}\cdot\ell\cdot V_{p-1},
$$
from which we have
$$
\vol (\Delta_p\diamondplus C)=\frac {q!\sqrt{p+1}}{(p+q)!}\,\cdot \left(\frac{\ell}{\sqrt 2}\right)^p\cdot\vol (C).
$$
%This recovers the well known fact that the volume of the regular $p$ simplex with the unit side length $\ell$
%is  given by $\frac {\sqrt{p+1}}{\sqrt {2^p}\, p!}\, \ell^p$.

With this formula, we have the following volumes:
$$
\begin{aligned}
\vol(\ghz_n)&=\tfrac {\sqrt{d}}{(d-1)!},\\
\vol(\fbi_n)&=\vol(\fbi_n^\triangle \diamondplus \fbi_n^\square)
   =\tfrac{(d/2)! \sqrt{d}}{(d-1)!} \left(\tfrac 2d\right)^{d/2}.
\end{aligned}
$$
Because $\gen$ consists of $d$ pieces of simplices with the the side length $1/\sqrt 2$, we also have
$$
\vol(\gen)=\tfrac {d\sqrt{d}}{{2^{d-1}}(d-1)!}.
$$
Therefore, we have the following relative volumes with respect to the whole simplex $\ghz_n$:
$$
\frac{\vol(\gen)}{\vol(\ghz_n)}=d\cdot \left(\frac 12\right)^{d-1},\quad % \to 0,\qquad
\frac{\vol(\fbi_n)}{\vol(\ghz_n)}
={(d/2)! \over (d/2)^{d/2}}% \le {1 \over d/2}\longrightarrow 0.
$$
Both of them tend to zero, as the number of qubits tends to infinity.

The volume radius of a set $X$ is given by the radius of a Euclidean ball whose volume is same as that of $X$,
as it was introduced in \cite{szarek_05}.
For subsets $X=\gen$, $\bisep_n\setminus\fbi_n$ and $\fbi_n$
of $\ghz_n$, we will consider the {\sl relative volume radius}
${\text{\rm rvr}}_n(X):=\left(\frac{\vol(X)}{\vol(\ghz_n)}\right)^{\frac 1{\dim}}$
with respect to the whole simplex $\ghz_n$. We have
$$
\begin{aligned}
{\text{\rm rvr}}_n(\ghz_n\setminus\bisep_n)&=\frac 12\, {d^{1/{(d-1)}}},\\
{\text{\rm rvr}}_n(\bisep_n\setminus\fbi_n)&=\left(1-\frac d{2^{d-1}}-\frac{(d/2)!}{(d/2)^{d/2}}\right)^{1/(d-1)},\\
{\text{\rm rvr}}_n(\fbi_n)&=\left(\frac{(d/2)!}{(d/2)^{d/2}}\right)^{1/(d-1)},
\end{aligned}
$$
and they approach $\frac 12$, $1$, $\frac 1{\sqrt e}$, respectively, as $n\to\infty$.
The last follows from
$$
{1 \over d-1} \log {(d/2)! \over (d/2)^{d/2}} = {1 \over d-1} \sum_{k=1}^{d/2} \log {k \over d/2}
\to {1 \over 2} \int_0^1 \log x dx = -{1 \over 2}.
$$

Now, we also consider the largest balls inside the polytopes $\ghz_n$, $\bisep_n$ and $\fbi_n$.
The largest ball inside density matrices, separable states and biseparable states have been considered
by several authors \cite{{harriman},{zyczk_98},{vidal_99},{braunstein_99},{gurvits_02},{gurvits_03},{jungnitsch_11}}.
%For an arbitrary real coefficients $(p_\bfi)_{\bfi\in I_n}$ with $\sum_\bfi p_\bfi=1$, we see that $\varrho_p$ is a density matrix
%if and only if
%$\left(\begin{matrix}p_\bfi+p_{\bar\bfi} &|p_\bfi- p_{\bar\bfi}|\\|p_\bfi- p_{\bar\bfi}|&p_\bfi+p_{\bar\bfi}\end{matrix}\right)$
%is positive (semi-definite) for every $\bfi\in I_n$ if and only if $p_\bfi\ge 0$ for every $\bfi\in I_n$.
%Therefore, $\varrho_p$ is a state if and only if $p$ is a probability distribution.
For a given fixed state $\varrho_p$ in $\ghz_n$, the radius $r_p$ of the
largest ball inside $\ghz_n$ around $\varrho_p$ is given by the
minimum distance from $\varrho_p$ to facets. The distance from
$\varrho_p$ to the facet $\ghz_n^\bfi$ can be obtained by the
distance to the linear manifold given by $\sum_\bfj p_\bfj=1$ and
$p_\bfi=0$. Using the Lagrange method, the distance is given by
$p_\bfi\sqrt{d/(d-1)}$ whose minimum over $\bfi\in I_n$ is just
$r_p$. Therefore, the maximum of $r_p$ occurs when $p$ is the
uniform distribution, and so we conclude that the largest ball
inside $\ghz_n$ is centered at the maximally mixed state and the
radius is given by $\sqrt{1/d(d-1)}$ which is the distance between
$c$ and $c^\bfi$ in {\sc Fig 1}. This number was shown in
\cite{harriman} to be the radius of the largest ball in the density
matrices. Our result shows that the maximum radius also occurs
within GHZ diagonal states. The exactly same argument shows that the
largest ball inside the polytope $\bisep_n$ coincides with the
largest ball inside $\ghz_n$. In order to find the largest ball
inside the polytope $\fbi_n$, we first compute the distance from a
state $\varrho_p$ to the facet $\fbi_n^{\bfi,\bfj}$, the linear
manifold given by $\sum_\bfk p_\bfk=1$ and (\ref{cccc}). If
$\bfj\in\{\bfi,\bar\bfi\}$ then the distance is given by
$p_{\bar\bfj}\sqrt{d/(d-1)}$ as before. Otherwise, we use the Lagrange
method again to get the distance $\frac 12|p_\bfi+p_{\bar\bfi}- p_\bfj +p_{\bar\bfj}|
\sqrt{d/(d-1)}$. From this, we conclude that the
largest ball inside $\fbi_n$ coincides again with the largest ball
inside the whole simplex $\ghz_n$.

Finally, we consider the convex set $\mermin_n$ of all $n$-qubit GHZ diagonal states which violate the Mermin
inequality. Because the hyperplane $H_M$ meet edges at the points in (\ref{points_mermin}),
we see that the volume of $\mermin_n$ is given by
$$
\vol(\mermin_n) = \vol(\ghz_n) \times ({1-\nu_n \over 2}) \times (1-\nu_n)^{d-2} = {(1-\nu_n)^{d-1} \sqrt{d} \over 2(d-1)!}.
$$
Note that the relative volume
$$
{\vol(\mermin_n) \over \vol(\ghz_n)} = {(1-\nu_n)^{d-1}\over 2}
$$
converges to zero, even though the vertices $w_\bfi$ of $\mermin_n$ converge to
$v_\bfi$ for $\bfi\neq {\bf 0}, {\bf 1}$
and to $m_{{\bf 0},{\bf 1}}$ for $\bfi=\bf 1$.
We see that the relative volume radius ${\text{\rm rvr}}_n(\mermin_n)=\frac 1{2^{1/(d-1)}}(1-\nu_n)$
converges to $1$ as $n\to\infty$.

\section{Conclusion}

In this paper, we have explored polytope structures for genuine entanglement,
biseparability, full biseparability and Bell inequality of multi-qubit GHZ diagonal states.
Through the discussion, we may visualize which parts of the simplices of all GHZ diagonal states represent genuine entanglement,
PPT states and those violating multipartite Bell inequality, respectively.
With these pictures, we have computed precise volume related values and their asymptotic behaviors
for genuine entanglement, biseparability, full biseparability or equivalently PPT, and violating Bell inequality.
All of them look reasonable, but we could not explain
why $\lim_{n\to\infty}{\text{\rm rvr}}_n(\fbi_n)$ is given by the number $\frac 1{\sqrt e}$.
We also have seen that the largest balls inside
three polytopes coincide. This means that the largest balls do not explain the relative volumes
in case of GHZ diagonal states.
It would be nice to compute the precise volume of the convex set consisting of fully separable GHZ diagonal states.
But this job must be much more involved, because fully separable GHZ diagonal states do not make a polytope anymore.
See \cite{{han_kye_GHZ},{chen_han_kye}}.

The authors are grateful to Hyun Kwang Kim for fruitful discussions on hypersimplices.
They are also grateful to the referee for bringing their attention to multipartite Bell inequality.
Both KHH and SHK were partially supported by NRF-2020R1A2C1A01004587, Korea.

\end{document}